\documentstyle[12pt,wrapfig]{article}
\renewcommand{\thefootnote}{\dagger}
\newcommand{\od}{\stackrel{\cdot}}
\newcommand{\td}{\stackrel{\cdot\cdot}}
\newcommand{\fd}{\stackrel{\cdot\cdot\cdot}}
\begin{document}
\begin{center}

{\Large
Characteristics of synchrotron radiation of longitudinally
polarized spinning particle in the pseudoclassical theory}
\vspace{1cm}

{
 Grigoryan G.V.\raisebox{.8ex}{$\star$}, Grigoryan R.P.
\raisebox{.8ex}{$\star\star$}}.
\footnote{ Partially supported  by
the grants INTAS 96-538,  INTAS 93-1038 and INTAS-RFBR 95-0829.}\\

{\em Yerevan Physics Institute, Republic of Armenia}\\

\vspace{1cm}
\raisebox{.8ex}{$\star$}{E-mail:gagri@lx2.yerphi.am}

\raisebox{.8ex}{$\star\star$}{E-mail:rogri@lx2.yerphi.am}\\
\end{center}

\centerline{{\bf{Abstract}}}
Spin dependencies of the intensity and polarization of the
synchrotron radiation of the longitudinally polarized particle
performing uniform circular motion in the magnetic field are
investigated.\\


\renewcommand{\thefootnote}{\arabic{footnote}}
\setcounter{footnote}{0}

As it is well known, the synchrotron radiation (SR) of the polarized
particle contains a contribution due to the spin.  Quantum
mechanical theory of the spin dependence of the
characteristics of the SR from polarized electron was developed
by Sokolov and Ternov \cite{STSY}. In \cite{NG} was suggested,
that the spin dependence of the SR may be used to measure the
degree of the beam transverse polarization.
In \cite{AAK} a method to measure the degree of the beam
longitudinal polarization using SR was proposed. All
estimations in these papers were based on the quantum
electrodynamics . 

Note, that as the spin of the particle is of the order of the
Planck constant $\hbar$, the contribution of the spin in the
characteristics of the SR to the first order of $\hbar$ is of
classical nature.  The semiclassical theory of the SR was
developed in \cite{TB}, however they did not consider the case of
longitudinal polarization of the electrons.

On the other hand, it is well known, that the spin of the
electron can be described  by Grassmann variables already at the
"classical" level in pseudoclassical theory of relativistic
spinning particle (after quantization this theory results in a
Dirac theory of the electron) The consistent description of the
spin in external fields using Grassmann variables already at the
classical level allows to investigate spin effects in radiation
processes using classical equations of motion.

Lienard-Wiechert potentials for the relativistic spinning
particle with anomalous magnetic moment in the pseudoclassical
approach were constructed in \cite{GG10}. In the latter the
general expressions for the Lienard-Wiechert potentials were used
for investigation of some specific cases of the motion of the
spinning particle. In particular, the spin  dependencies of
intensity and the polarization of the SR  of
the transversely polarized particle performing uniform circular
motion were investigated. It is important to stress,  that within
our approach the evaluation of the formulae,  describing the spin
dependence of the measurable characteristics of the relativistic
particle radiation is much easier,  than in quantum
electrodynamics.

In this paper general expressions obtained in \cite{GG10} will be
used for investigation of the spin dependencies of the intensity
and polarization of the SR of the
longitudinally polarized particle performing uniform circular
motion in the magnetic field.\\

Pseudoclassical theory of the interaction of the relativistic
spinning particle with anomalous magnetic moment (AMM) with
electromagnetic field is a sample of the constrained theories.
Acting in the standard way, when additional constraints are
introduced into the theory for complete fixation of the gauges
and keeping only terms up to second order in Grassmann variables
(which is equivalent to  keeping only terms, which are of the
first order in spin), we obtain for the potentials of the
electromagnetic field $A_\mu$ the equation (for details see
\cite{GG10})
\begin{equation}
\label{L1}
\Box A^\mu=j^\mu,
\end{equation}
where $j^\mu$ is given by the expression
\begin{equation}
\label{L2}
j^\mu(y)=g\int d\tau\dot x^\mu\delta(x(\tau)-y)+
\frac{\partial}{\partial y^\nu}\int
d\tau\delta(x(\tau)-y)p^{\nu\mu}(\tau).
\end{equation}
Here $y$ is observer's coordinate, $g$ is the charge, $x_\mu$
-the coordinate of the  particle, the overdote denotes the
differentiation over $\tau$ along the  trajectory, $p_{\mu\nu}$
is the tensor of the dipole moment of the particle.

As it was done in \cite{GG10}, here also the physical constraints
fixing the gauges of the theory will be chosen in the form
\begin{equation}
\label{L2A}
\partial_\mu A^\mu=0,\quad \od{x}^2=1,\quad
q_\mu=p_{\mu\nu}\od{x}^\nu=0.
\end{equation}
The second relation in (\ref{L2A}) means that the parameter
$\tau$ becomes a proper time of the particle, while the third
relations corresponds to the fact, that the electric dipole
moment of the particle $q_\mu$ is equal to zero, as it should be
for a point particle.

The first summand in (\ref{L2}) corresponds to the current of the
particle without a dipole moment and the second term corresponds
to the contribution to the current  of  the dipole moment  of the
spinning particle.

In the chosen gauge for the theory of the particle without AMM
considered here
 $p_{\mu\nu}$ is given by the expression \cite{GG10}:
\begin{equation}
\label{L3}
p_{\mu\nu}=\frac{g}{m^2}
\varepsilon_{\mu\nu\lambda\sigma}W^\lambda\od {x}{\!}^\sigma.
\end{equation}
Here $W^\mu$ denotes the pseudoclassical analog of the Pauli-
Lubansky vector, which is connected with the vector of the
particle spin in the particle rest frame by the relation
\begin{equation}
\label{L4}
\frac{W_\mu}{m}
=\left(\gamma(\vec{v}\vec{S}),\,\,{S}_i+
\gamma^2\frac{{v}_i(\vec{v}\vec{S})}{\gamma+1}\right),
\end{equation}
where $\vec{v}\equiv(v^i)=dx^i/dt$ is three dimensional velocity
of the particle, $dt/d\tau=\gamma=\left(\sqrt{1-v^2}\right)^{-1}$
(in pseudoclassical theory the tensor $p_{\mu\nu}$ and vectors
$W_\lambda$ and  $S_i$ are quadratic in Grassmann variables;
their explicit expressions are irrelevant for this investigation).

The solution of the equation   (\ref{L1}) (with the current
represented in the form of (\ref{L2})) in terms of the retarded
fields was found in \cite{EL}. Here we will present only that
part of the general expression, which corresponds to the
particle radiation field:
\begin{equation}
\label{L5}
F_{{\rm ret}}^{\mu\nu}(y)=\frac{1}{2\pi\rho}
\left[g\left(k^{[\mu}\td{x}{\!}^{\nu]}-
(\td{x}k)k^{[\mu}\od{x}{\!}^{\nu]}\right)
+P_1^{\left[\mu\nu\right]}
\right]_{\tau=\tau_r},\nonumber
\end{equation}
where $\rho = \od{x}{\!\!}^\nu R_\nu$, $k_\mu=R_\mu/\rho$,
$R^\mu\equiv y^\mu-x^\mu(\tau) $, $R^\mu R_\mu=0$; all quantities
in  (\ref{L5}) are evaluated at the retarded time
 $\tau_r$ defined by the equation $t_0=t_0(\tau_r)=t-R(t_0)$
and  $P_1^{\mu\nu}$ is given by the
expression:
\footnote{Here and below square brackets
$[ \dots ]$ denote complete antisymmetrization:e.g.,

\centerline{$A_{[\alpha\beta\gamma]}=1/3!\left\{
A_{\alpha\beta\gamma}+A_{\beta\gamma\alpha}+A_{\gamma\alpha\beta}-
A_{\beta\alpha\gamma}-A_{\alpha\gamma\beta}-A_{\gamma\beta\alpha}\right\};$}
round brackets $(\dots)$ denote complete symmetrization: e.g.,

\centerline{$A_{(\alpha\beta\gamma)}=1/3!
\left\{A_{\alpha\beta\gamma}+A_{\beta\gamma\alpha}+A_{\gamma\alpha\beta}+
A_{\beta\alpha\gamma}+A_{\alpha\gamma\beta}+A_{\gamma\beta\alpha}\right\}.$}}.
\begin{equation}
\label{L6}
P_1^{\mu\nu}=T^{\mu\alpha}k_\alpha k^\nu,\quad
T^{\mu\nu}=\td{p}{\!}^{\mu\nu}
-3(\td{x}k)\od{p}{\!}^{\mu\nu}+
3(\td{x}k)^2{p}^{\mu\nu}-
(\fd{x}k){p}^{\mu\nu}.
\end{equation}

In formulae  (\ref{L5}) the first term corresponds to the contribution
 of the charge of the particle,  the second -to the contribution of the
dipole moment.

Consider now the case of the longitudinally polarized particle
($\vec{v}||\vec{S}$) moving in the homogeneous magnetic field
$\vec{B}$. As it is well known, the angular velocity of the
particle without anomalous magnetic moment in this case is equal
to the spin precession angular velocity and the chirality of the
particle doesn't change during its motion. Then we have the
relation
\begin{eqnarray}
\label{L7}
&&v^k=\lambda\frac{v}{S}S^k,\quad \lambda=\pm 1,\\
&&(\vec{v}\vec{a})=(\vec{S}\vec{a})=0,\quad  \od{\gamma}=0,\quad
\od{\vec{\omega}}=0,
\end{eqnarray}
where $\vec{a}=d\vec{v}/dt$.

Taking into account these relations and also the expressions
(\ref{L4}) we obtain for the  components $p_{0i}$ and
$p_{ij}$ of the dipole moment tensor (\ref{L3}) and their
derivatives relations
\begin{equation}
\label{L8}
p_{0i}=\od{p}_{0i}=\td{p}_{0i}=0,
\end{equation}
\begin{eqnarray}
\label{L9}
p_{ij}=-\frac{g}{m}\varepsilon_{ijk}S_k,&& \quad
\od{p}{}_{ij}=-\frac{g}{m}\varepsilon_{ijk}\od{S}{}_k=-
\gamma\frac{g}{m}\left(\omega_iS_j-\omega_jS_i\right)),\nonumber\\
&&\td{p}{}_{ij}=\gamma^2\frac{g}{m}\omega^2\varepsilon_{ijk}S_k.
\end{eqnarray}

Returning now to the formula (\ref{L5}) we find for the
contribution of the dipole moment of the particle to the particle
radiation field (denoted  by $E_i^{\rm dip}$):
\begin{eqnarray}
\label{L10}
E_i^{{\rm
dip}}&=&-\left(F_{0i}^{\rm rad}\right)^{\rm dip}=
-\frac{1}{4\pi\rho}\left(P_{1,0i}
-P_{1,i0}\right)=\nonumber\\\
&=&-\frac{1}{4 \pi\rho}\left[T_{0l}k^lk_i-T_{i\lambda}k^\lambda
k_0\right]_{\tau=\tau_r}=
-\frac{1}{4
\pi\rho}\left[\left(\delta_{ij}k_0^2-k_ik_j\right)T_{0j}
+T_{ij}k_jk_0\right]_{\tau=\tau_r},
\end{eqnarray}
where $T_{0i}$ and $T_{ij}$ are components of the tensor $T_{\mu\nu}$.
It is easy to see that the components $T_{0i}$ are identically
zero due to the relations (\ref{L8}). Now taking into account
the relations
\begin{eqnarray}
\label{L11}
\od{x}{\!}^\mu&=&\gamma(1,
\vec{v}), \quad\rho=\gamma(R-(\vec{v}\vec{R}))=\gamma
R(1-(\vec{v}\vec{n})), \quad  \gamma=1/\sqrt{1-v^2},\nonumber\\
\td{x}{\!}^\mu&=&\left(\gamma^4(\vec{v}\vec{a}), \, \, \gamma^2\vec{a}+
\gamma^4\vec{v}(\vec{v}\vec{a})\right), \quad
k^\alpha=\left(\frac{R}{\rho}, \frac{\vec{R}}{\rho}\right)=
\frac{R}{\rho}n^\alpha,\quad n^\alpha=\frac{R^\alpha}{R}, \\
\fd{x}{\!}^\mu&=&\left(\od{a^0}, \, \, \gamma^3{d\vec{a}/dt}+
3\gamma^5 (\vec{v}\vec{a})\vec{a}+\od{a^0}\vec{v}\right), \quad
\od{a^0}=\gamma^5\left[(\vec{v}d\vec{a}/dt)+\vec{a}^2\right]+
4\gamma^7(\vec{v}\vec{a})^2, \nonumber
\end{eqnarray}
we transform the formula (\ref{L10}) into
\begin{eqnarray}
\label{L12}
E_i^{{\rm dip}}&=&
\frac{R^2}{4\pi\rho^3}T_{ij}n_j=
\frac{1}{4\pi R(1-(\vec{n}\vec{v}))^3}\frac{g}{m\gamma}
\left[\frac{3(\vec{a}\vec{n})}
{1-(\vec{n}\vec{v})}\left(\omega_i(\vec{S}\vec{n})
-S_i(\vec{\omega}\vec{n})\right)+\right.\nonumber\\
&+&\left.\varepsilon_{ijk}n_jS_k\left(\frac{3(\vec{a}\vec{n})^2}
{(1-(\vec{n}\vec{v}))^2}-\frac{\omega^2}{1-(\vec{n}\vec{v})}\right)
\right]_{\tau=\tau_r}.
\end{eqnarray}

Substituting expressions (\ref{L11}) into first term of the
formula (\ref{L5}) we get the well known expression for the
strength of the electric field $E_i^{\rm ch}$ of the spinless
charged particle radiation:
\begin{eqnarray}
\label{L13}
E_{i}^{{\rm ch}}&=&\frac{g}{4\pi R(1-(\vec{v}\vec{n}))^3}
\left[\left(n_i-v_i\right)(\vec{a}\vec{n}) -
a_i(1-(\vec{v}\vec{n}))\right]_{\tau=\tau_r},\\
E_i^{\rm rad}&=&-F_{0i}^{\rm rad}=E_i^{\rm ch}+E_{i}^{\rm
dip}.\nonumber
\end{eqnarray}

The intensity of the   SR is known to be
proportional to $(E_i^{\rm rad})^2=\left(E_i^{\rm ch}+E_{i}^{\rm
dip}\right)^2$. The contribution to the radiation intensity
of the dipole moment of the particle to the first order
in spin  evidently will be defined by the term $ 2E_i^{\rm
ch}E_{i}^{\rm dip}$. Using expressions
(\ref{L12}) and (\ref{L13}), we find
\begin{eqnarray}
\label{L14}
2E_i^{{\rm ch}}E_i^{{\rm
dip}}&=&2\left(\frac{g}{4R\pi}\right)^2\frac{\lambda S v}
{m\gamma(1-(\vec{v}\vec{n}))^6} \omega^2
(\vec{\omega}\vec{n})=\nonumber\\
&=&-2\left(\frac{g}{4R\pi}\right)^2\frac{\lambda S v\omega^3}
{m\gamma(1-v\cos\theta)^6}
\sin\theta\sin\varphi,
\end{eqnarray}

\begin{wrapfigure}[11]{l}[8pt]{5cm}
\unitlength=1mm
\begin{picture}(60,60)
\put(20,25){\vector(1,0){30}}
\put(20,25){\vector(0,1){30}}
\put(20,25){\vector(-1,-1){20}}
\put(20,25){\vector(1,1){20}}
\put(20,25){\vector(1,2){10}}
\multiput(20,25)(5,-5){3}{\line(1,-1){4}}
\multiput(30,15)(0,3){10}{\line(0,1){2}}
\put(20,35){\oval(10,10)[tr]}
\put(26,25){\oval(12,12)[br]}
\put(4,5){\shortstack{$\vec{B}$}}
\put(23,41){\shortstack{$\theta$}}
\put(12,52){\shortstack{$\vec{v}(\vec{S})$}}
\put(36,37){\shortstack{$\vec{\omega}$}}
\put(46,27){\shortstack{$\vec{a}$}}
\put(31,43){\shortstack{$\vec{n}$}}
\put(33,21){\shortstack{$\varphi$}}
\put(23,5){\shortstack Fig.1}
\end{picture}
\end{wrapfigure}

where $\theta$ and $\varphi$ are angles, defining the direction
of the vector $\vec{n}$ in the coordinate system, defined by the
vectors $\vec{v},\,\vec{a}$,  $\vec{B}$ (see Fig.1). 
Taking into account the expression (\ref{L14}) we write for the
contribution of the dipole moment to the intensity of the synchrotron
radiation in the first order of spin
\begin{equation}
\label{L15}
\frac{d{\cal I}^{dip}}{dt_0 d\Omega}=-\left(\frac{g}{4\pi}\right)^2
\frac{2\lambda}{(1-v\cos\theta)^5}
\frac{\omega^3 vS\sin\theta\sin \varphi}{m\gamma}.
\end{equation}
From (\ref{L15})
one can see that the integral contribution of the
dipole moment of the particle to the
intensity of the SR in the first order of
spin for longitudinally polarized particle is equal to zero.

Integrating (\ref{L15}) over the upper hemisphere
($0\leq\varphi\leq
\pi$,\,$0\leq\theta\leq \pi$), we get
\begin{equation}
\label{L16}
\frac{d{\cal I}^{dip}_{{\rm up}}}{dt_0}=-
\frac{g^2}{4\pi}\frac{4+v^2}{8m}\omega^3v\lambda S\gamma^6.
\end{equation}
From (\ref{L15}), taking into account, that
\begin{equation}
\label{L17}
\frac{d{\cal I}^{dip}_{{\rm up}}}{dt_0}=-\frac{d{\cal I}^{dip}_{{\rm
down}}}{dt_0},
\end{equation}
and that
\begin{equation}
\label{L17A}
\frac{d {\cal I}_{up(down)}^{tot}}{dt_0}=\frac{g^2}{4\pi}\frac{a^2}{3}\gamma^4+
\frac{d{\cal I}^{dip}_{{\rm up(down)}}}{dt_0}
\end{equation}
we find for the asymmetry of the SR into upper
(up) and down (down) hemispheres the expression
\begin{equation}
\label{L18}
\delta=\frac{\displaystyle\frac{d{\cal I}^{dip}_{{\rm up}}}{dt_0}-
\frac{\displaystyle d{\cal I}^{dip}_{{\rm
down}}}{dt_0}}{\displaystyle\frac{d{\cal I}^{tot}_{{\rm
up}}}{dt_0}+
\frac{\displaystyle d{\cal I}^{tot}_{{\rm
down}}}{dt_0}}=-\frac{3(4+v^2)}{8mv}\omega\gamma^2\lambda S.
\end{equation}
In ultrarelativistic case we have
\begin{equation}
\label{L19}
\delta=-\frac{15}{8m}\omega\gamma^2\lambda S.
\end{equation}

Now taking into account that $S=1/2$, the relation $\omega m\gamma=gB$ and
introducing the Shwinger field $H_0=m^2/g=4.41\cdot
10^{9}T$, we can rewrite the last equation in the form
\begin{equation}
\label{L20}
\delta=-\frac{15}{16}\gamma\lambda\frac{B}{H_0}.
\end{equation}
For LEP energies of $E=45 Gev$ wiggler magnetic fields $B=2T$ we
find for $\lambda=1$ $\delta=-0,38\cdot 10^{-4}$, which coincides with the
calculations based on the quantum electrodynamics \cite{AAK}

\begin{wrapfigure}[12]{r}[8pt]{6cm}
\unitlength=1mm
\linethickness{0.2mm}
\begin{picture}(60,56)
\put(20,25){\vector(1,0){30}}
\put(20,25){\vector(0,1){30}}
\put(20,25){\vector(-1,-1){20}}
\put(20,25){\vector(1,2){10}}
\multiput(20,25)(0,-3){4}{\line(0,-1){2}}
\multiput(20,25)(5,-5){3}{\line(1,-1){4}}
\put(20,13){\vector(0,-1){4}}
\multiput(30,15)(0,3){10}{\line(0,1){2}}
\put(22,18){\oval(10,20)[tr]}
\put(30,25){\oval(20,20)[br]}
\put(4,5){\shortstack{$\vec{a}$}}
\put(26,29){\shortstack{$\beta$}}
\put(16,52){\shortstack{$\vec{B}$}}
\put(21,11){\shortstack{$\vec{\omega}$}}
\put(46,27){\shortstack{$\vec{v}$}}
\put(31,43){\shortstack{$\vec{n}$}}
\put(40,16){\shortstack{$\psi$}}
\put(23,3){\shortstack{Fig.2}}
\end{picture}
\end{wrapfigure}

To investigate the characteristics of the polarization of
SR we resolve (following \cite{BKF}) the electric field
strength $\vec{E}^{\rm rad}$ into components along the vectors
$\vec{e}=\vec{a}/a$  and  $\left[\vec{n}, \vec{e}\right]$:
\begin{equation}
\label{L21}
\vec{E}^{\rm rad}=E_1 \vec{e}+E_2\left[\vec{n}, \vec{e}\right].
\end{equation}
This choice of unit vectors is suitable for description of
the radiation of the relativistic particles using angles defined
on Fig.2,  which are appropriate since the main contribution to
the radiation comes from small $(\sim 1/\gamma)$ angles $\beta$
and $\psi$,  in accordance with the fact,  that the angle between
vectors $\vec{n}$ and $\vec{v}$ is of the order of $(\sim
1/\gamma)$. As it was mentioned in \cite{BKF},  the vector
$\vec{E}$ is not orthogonal to vector $\vec{n}$,  however this
deviation is of the order of $1/\gamma$,  so that the
decomposition (\ref{L21}) can be used for calculation of major
terms (with accuracy $1/\gamma$).
From formulae (\ref{L12}) and  (\ref{L13}) we get
according to decomposition (\ref{L21}) the expressions for
$E_1$  and $E_2$:
\begin{equation}
\label{L22}
E_1=\frac{ga}{\pi
R(\mu^2+\psi^2)^3}\left((\psi^2-\mu^2)+\frac{4a\lambda S}{m\gamma}
\frac{\beta(\mu^2-5\psi^2)}{(\mu^2+\psi^2)^2}
\right)_{\tau=\tau_r},
\end{equation}
\begin{equation}
\label{L23}
E_2=-\frac{2ga}{\pi
R(\mu^2+\psi^2)^3}\left(\beta\psi-\frac{4a\lambda S}{m\gamma}
\frac{\psi(2\mu^2-\psi^2)}{(\mu^2+\psi^2)^2}
\right)_{\tau=\tau_r},
\end{equation}
where $\mu^2=\gamma^{-2}+\beta^2$.

Authors are thankful to I.V.Tyutin and A.Airapetian for useful
discussion of the topics of the paper.

This investigation was supported in part by the grants
INTAS 96-538,  INTAS 93-1038 ¨ INTAS-RFBR 95-0829.

\end{document}